\documentclass[prl,twocolumn,reprint,showpacs,preprintnumbers,amsmath,amsthm,
amssymb,amsfont]{revtex4-1}
\usepackage[version=4]{mhchem}
\usepackage{times}
\usepackage{graphicx}
\usepackage{dcolumn}
\usepackage{bm}
\usepackage{color}
\usepackage{subfigure}

\begin{document}

\title{Giant Ferrimagnetism
and Polarization In A Mixed Metal Perovskite Metal-Organic Framework }
\author
{Paresh C. Rout$^{(1)}$and Varadharajan Srinivasan$^{(1,2)}$}
\affiliation{(1) Department of Physics, Indian Institute of Science Education
and Research Bhopal, Bhopal 462 066, India}
\affiliation{(2) Department of Chemistry, Indian Institute of Science Education
and Research Bhopal, Bhopal 462 066, India}

\begin{abstract}
Perovskite metal-organic frameworks (MOFs) have recently emerged as potential
candidates for multiferroicity. However, the compounds synthesized
so far possess only weak ferromagnetism and low polarization. Additionally, the
very low magnetic transition temperatures ($T_c$) also pose a challenge to the
application of the materials.We have computationally designed a mixed metal
perovskite MOF -\ce{[C(NH2)3][(Cu$_{0.5}$Mn$_{0.5}$)(HCOO)3]}- that is
predicted to have magnetization two orders of magnitude larger than its parent
(\ce{[C(NH2)3][Cu(HCOO)3]}), a significantly larger polarization (9.9
$\mu$C/cm$^2$), and an enhanced $T_c$ of up to 56 K, unprecedented in perovskite
MOFs. A detailed study of the magnetic interactions revealed a novel mechanism
leading to the large moments as well as the increase in the $T_c$. Mixing a 
non-Jahn-Teller ion (Mn$^{2+}$) into a Jahn-Teller host (Cu$^{2+}$) leads to
competing lattice distortions which are directly responsible for the enhanced
polarization. The MOF is thermodynamically stable as evidenced by the computed
enthalpy of formation, and can likely be synthesized. Our work represents a
first step towards rational design of multiferroic perovskite MOFs through the
largely unexlpored mixed metal approach.
\end{abstract}

\maketitle

Multiferroics are materials which possess ferromagnetic (FM), ferroelectric (FE)
and structural order parameters within a single
phase~\cite{Matter,Nature,Picozzi,Tokura,khom,Li,Ren,Lee}. These are highly
promising not only for their use in multi-functional device applications but
also for the interesting physics they reveal. Much of the research in the field
has so far focussed on multiferroics based on inorganic transition metal oxides.
In the last decade, there has been growing interest in metal-organic frameworks
(MOFs) consisting of metal ions interconnected by organic linkers. The
organic-inorganic duality in MOFs  leads to many interesting physical
properties~\cite{Rosse,Zhang} that can be exploited in applications such as gas
storage and separation, catalysis, nonlinear optics, photoluminescence,
magnetic and electric materials, and so on~\cite{Advanced,Appli}. The hybrid
nature of these materials offers a vast chemical space for synthetic chemists to
explore and, hence, also affords tunability of properties. MOFs with the
perovskite ABX$_{3}$ structure are of great interest, particularly  those with
multiferroic behavior arising due to hydrogen-bonds~\cite{Stroppa,Jain}. In the
case of magnetic MOFs, for instance, one can control the nature of magnetic
coupling through the variety of possible metal ions in the B-site,
short ligands, co-ligands and radical ligands carrying spin degrees of
freedom~\cite{Demir}. Recently, it has been shown that one can tune the
magnitude of the ferroelectric polarization by carefully choosing different
A-site cations in these MOFs~\cite{Sante}. 

In recent past, a new class of ABX$_{3}$ metal formates
[\ce{C(NH2)3][M(HCOO)3}] (abbreviated below as M-MOF, M= divalent Mn, Fe, Co,
Ni, Cu, and Zn), was experimentally synthesized~\cite{Hu}. Of these only the
Cu-MOF crystallizes into a polar space group ({\it Pna2}$_{1}$) and exhibits 
multiferroic and magnetoelectric behavior. It has been reported that the Cu-MOF
shows canted-spin anti-ferromagnetism with a N\'{e}el temperature of 4.6~K.
Using first-principles calculations, Stroppa {\it et al.}~\cite{Stroppa} showed 
that this polar Cu-MOF has a polarization of 0.37 $\mu$C/cm$^{2}$ along with a
weak magnetization. The polarization originates mainly from the displacements of
the A-site organic cation induced by hydrogen-bonds between the guanidinium
hydrogens and the oxygens of the formate linkers. The magnetization arises from
the transition metal (TM) ion at the B-site, where in-plane anti-ferro orbital
(AFO) ordering of Cu $d$-orbitals results in an anti-ferromagnetic (AFM) ground
state with A-type spin ordering. The low values of the polarization
and magnetization along with its low magnetic transition temperature ($T_c$)
precludes the Cu-MOF from being practically useful. These intrinsic drawbacks
can, in principle, be overcome by varying the A-site or B-site composition of
the MOF. A mixed metal strategy for the B-site ion (or B-site doping) has proven
to be successful in improving magnetic properties in inorganic multiferroic
compounds~\cite{Nicola,Spaldin}. The double-perovskites thus formed,
with TM ions of differing $d$-orbital configurations, can not only result in
larger magnetization but can also enhance the strength of the exchange coupling
interactions pushing the transition temperature higher. 
However, only a few studies have so far appeared that explore
this strategy~\cite{ciupa,gagor,mazzuca,zhaojp}. In particular, B-site doping in
perovskite MOFs aimed at improving ferroic properties is nascent~\cite{CuCd,
shang,Mazaka}. Moreover, to the best of our knowledge, there are no
theoretical predictions of mixed metal perovskite MOFs. First-principles 
based theory can not only help identify potential candidates but also elucidate
the key mechanisms driving ferroic orders in these MOFs. 

In this study, we have employed first-principles DFT-based techniques to
investigate the potential of mixed metal perovskite MOFs -
(M$_{0.5}$M$^\prime_{0.5}$)-MOF- as multiferroic materials. In particular, we
propose a novel mixed metal MOF- (Cu$_{0.5}$Mn$_{0.5}$)-MOF - which not only
yielded a magnetic moment two orders of magnitude larger than the parent Cu-MOF
but also a significantly larger transition temperature. The combination of
Mn$^{2+}$ and Cu$^{2+}$ was chosen deliberately keeping in mind the similarity
in sizes of the ions as well as the fact the pair represents the largest
different in magnetic moments possible on a ferrimagnetic lattice.
Indeed, the proposed MOF was found to have a magnetization of 4 $\mu_B$ per
Cu-Mn pair (or 2$\mu_B/$TM) which is the largest among mixed metal magnetic
MOFs synthesized so far. Since the parent Cu-MOF has a Jahn-Teller (JT) ion (Cu$^{2+}$), 
mixing in a non-JT ion (Mn$^{2+}$) would lead to competing lattice distortions which could
significantly influence the dielectric properties. Particularly, compositions in the vicinity of Cu$_{0.5}$Mn$_{0.5}$ are
expected to be more responsive as has recently been suggested~\cite{CuCd}. 
Surprisingly, the polarization in the compound was
significantly enhanced (9.9 $\mu$C/cm$^2$) compared to its parent. Furthermore,
doping with Mn$^{2+}$ ions resulted in an enhancement of the exchange coupling
between the TM ions.  This in turn increased the magnetic transition
temperatures to 24 K and 56 K, respectively, depending on the cation-ordering at
the B-site. The computed energy of formation indicates that the
(Cu$_{0.5}$Mn$_{0.5}$)-MOF is thermodynamically stable and, in principle, can be
synthesized. Our work highlights the potential of the largely unexplored mixed
metal strategy towards improving the ferroic properties of perovskite MOFs. 

Our spin-polarized DFT calculations employed a generalized gradient
approximation (GGA) to the exchange-correlation functional through the PBE
functional~\cite{pbe1,pbe2}. We accounted for correlation effects in the $3d$
TM ions through a DFT+$U$ approach~\cite{anisimov,matteo,wentzco}. 
We chose $U$ values of 3.5 and 4.0 eV for Mn and Cu, respectively, through a
self-consistent calculation of the parameter~\cite{marzari,heither,kulik}. 

In order to properly account for the weak interactions in the MOF, we have also 
incorporated a van der Waals' corrected functional~\cite{Grimme} in all our calculations. 
This GGA+$U$+vdW was used to perform structural
optimization calculations to obtain relative energies of magnetic and cation
orders, energies of formation, exchange coupling constants, etc. All calculations
were done using the plane-wave basis Quantum-ESPRESSO code~\cite{paulo} . 
All structures were fully optimized until forces were less than
0.26 meV/{\AA} on each atom. In the results presented below all bond lengths
are in angstrom (\AA) and energies are reported in meV per TM (meV/TM). The calculation methodology
was thoroughly tested for convergence of parameters and accuracy of the functionals employed
as detailed in the Supplementary Information (SI).

\begin{figure}[pbht!]
\begin{center}
\subfigure[]
{
\includegraphics[trim=0cm 0.0cm 0cm
2.0cm,clip=true, width=0.5\textwidth]{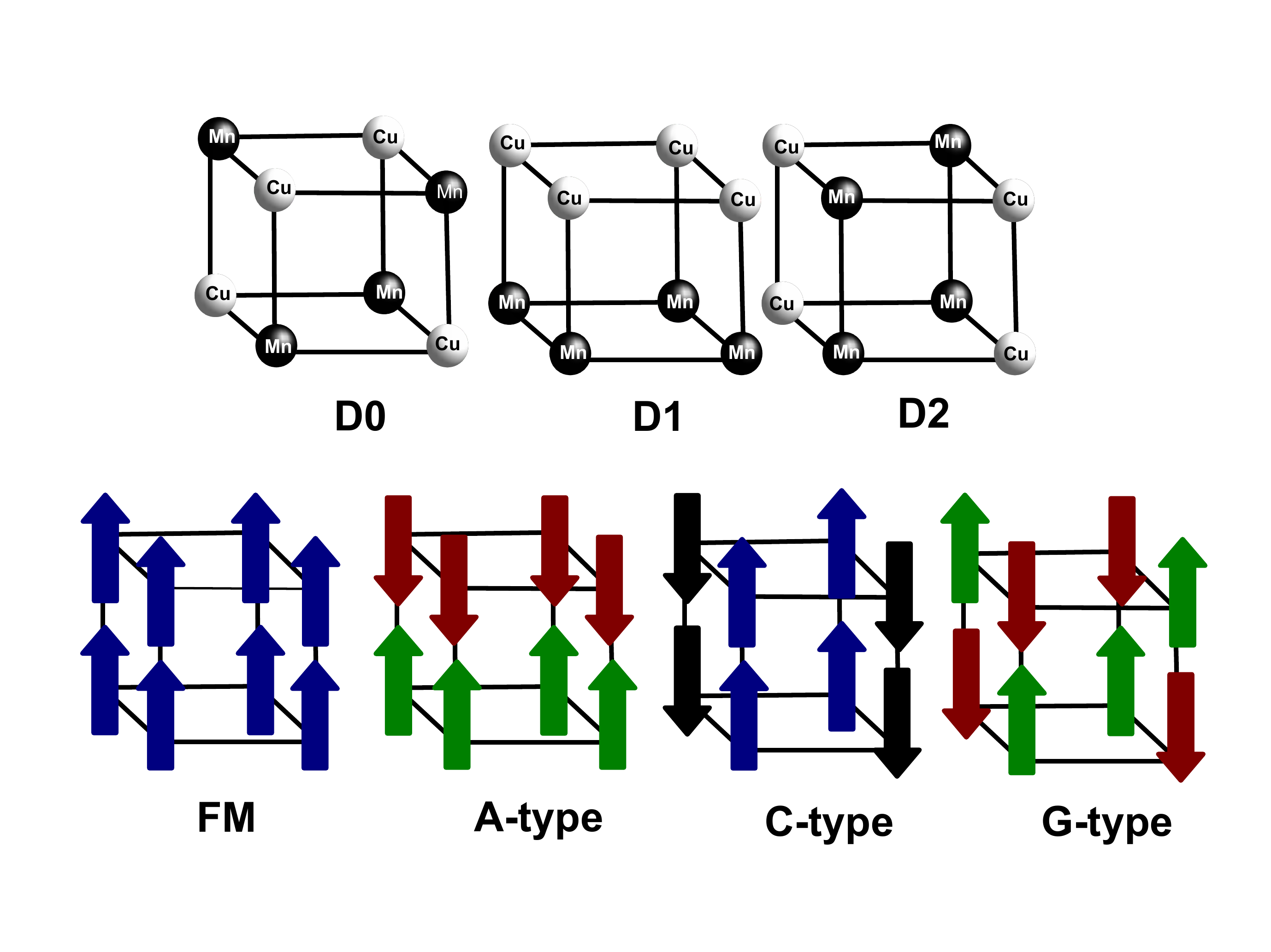}
}
\subfigure[]
{
\includegraphics[trim=0cm 0.0cm 0cm
2.0cm,clip=true, width=0.5\textwidth]{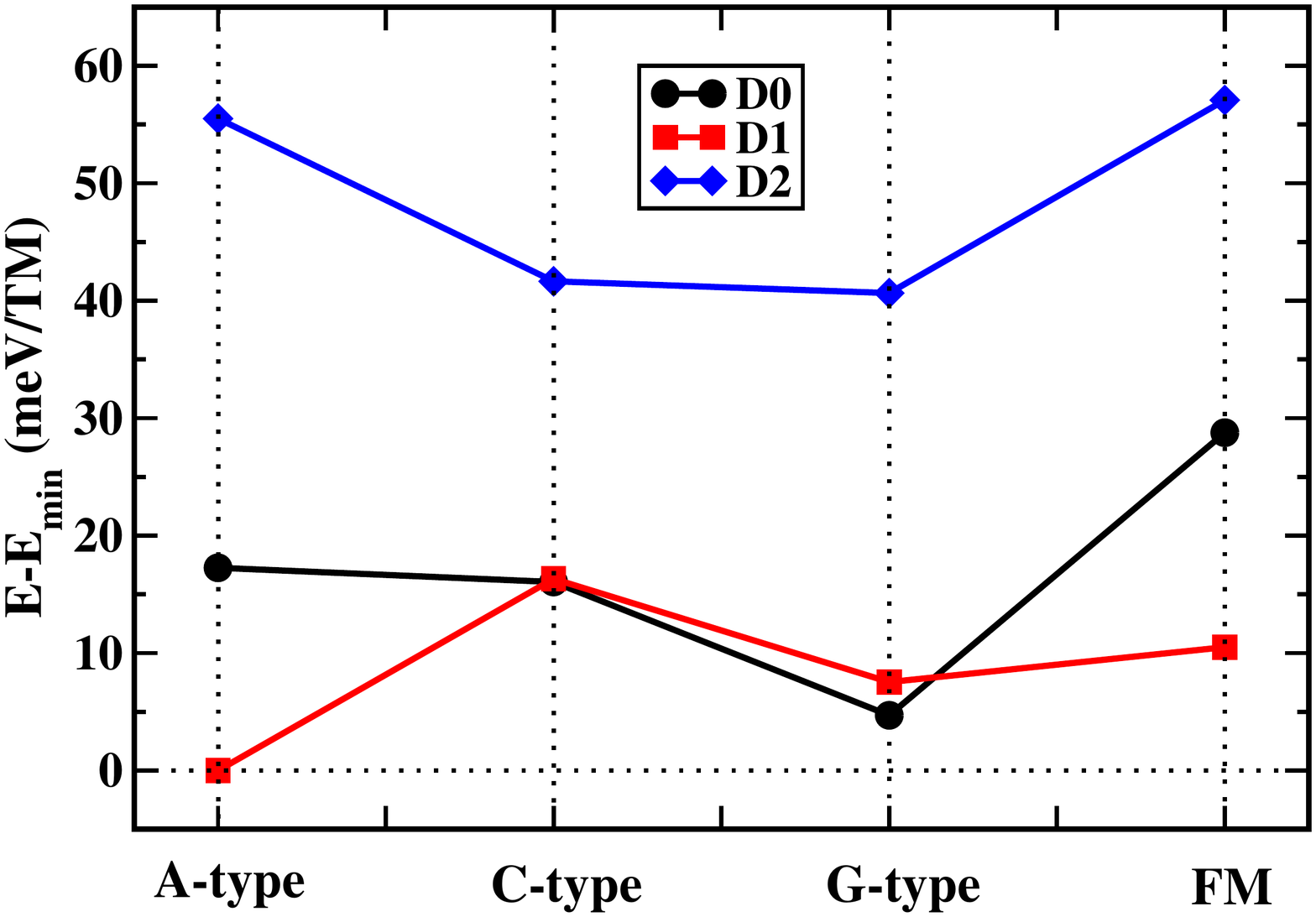}
}
\caption{Optimized energies of (Cu$_{0.5}$Mn$_{0.5}$)-MOF with all possible spin
orderings plotted with respect to various types of cation orderings. The following
symbols are used for different types of cation orders: circles for the D0,
squares for the D1, and diamonds for the D2 structures. A, C and G-type refer to
various antiferromagnetic spin orderings while FM refers to a ferromagnetic one.
\label{fig:2} }
\end{center}
\end{figure}

The unit cell of Cu-MOF contains four formula units. We produced the
(Cu$_{0.5}$Mn$_{0.5}$)-MOF by replacing two of the formula units by their Mn
analogues. This can result in three kinds of cation ordering (D0, D1 and D2)
as shown in Figure~\ref{fig:2}(a). For each cation
ordered structure we also investigated different collinear magnetic ordering of
the Mn (5 $\mu_B$) and Cu (1 $\mu_B$) spin moments (see Figure~\ref{fig:2}(a)). These included three AFM
arrangements (A, C, and G-type) and the ferromagnetic arrangement (FM). The optimized energies for the various structures considered
are summarized in the plot shown in Figure~\ref{fig:2}(b). The lowest energy
structure consists of layers of Mn and Cu alternating along the c-axis with an
A-type AFM arrangement of spins (referred to below as D1-A). A structure with
rock-salt ordering of the TM ions and with a G-type AFM arrangement of their
spins (referred to below as D0-G), was found to be higher in energy than D1-A by
just 4.75 meV/TM. Starting from the experimental structure of Mn-MOF
(\ce{[C(NH2)3][Mn(HCOO)3]}) ($Pnna$), we have  computed the ground
state of Mn-MOF to be G-type AFM. To the best of our knowledge, the ground
magnetic state of the MOF has not been reported earlier. We used the
predicted ground state of Mn-MOF for the calculation of formation
energy. The formation energies computed for the D1-A and D0-G
structures (-101 and -96.25 meV/TM, respectively) suggest that both can
likely be synthesized. We focus on these two structures as they are magnetic in
nature with moments comparable to inorganic compounds as shown below. 
\begin{figure}[pbht!]
\includegraphics[trim=0cm 0.0cm 0cm
1.0cm,clip=true,width=0.480\textwidth,clip=]{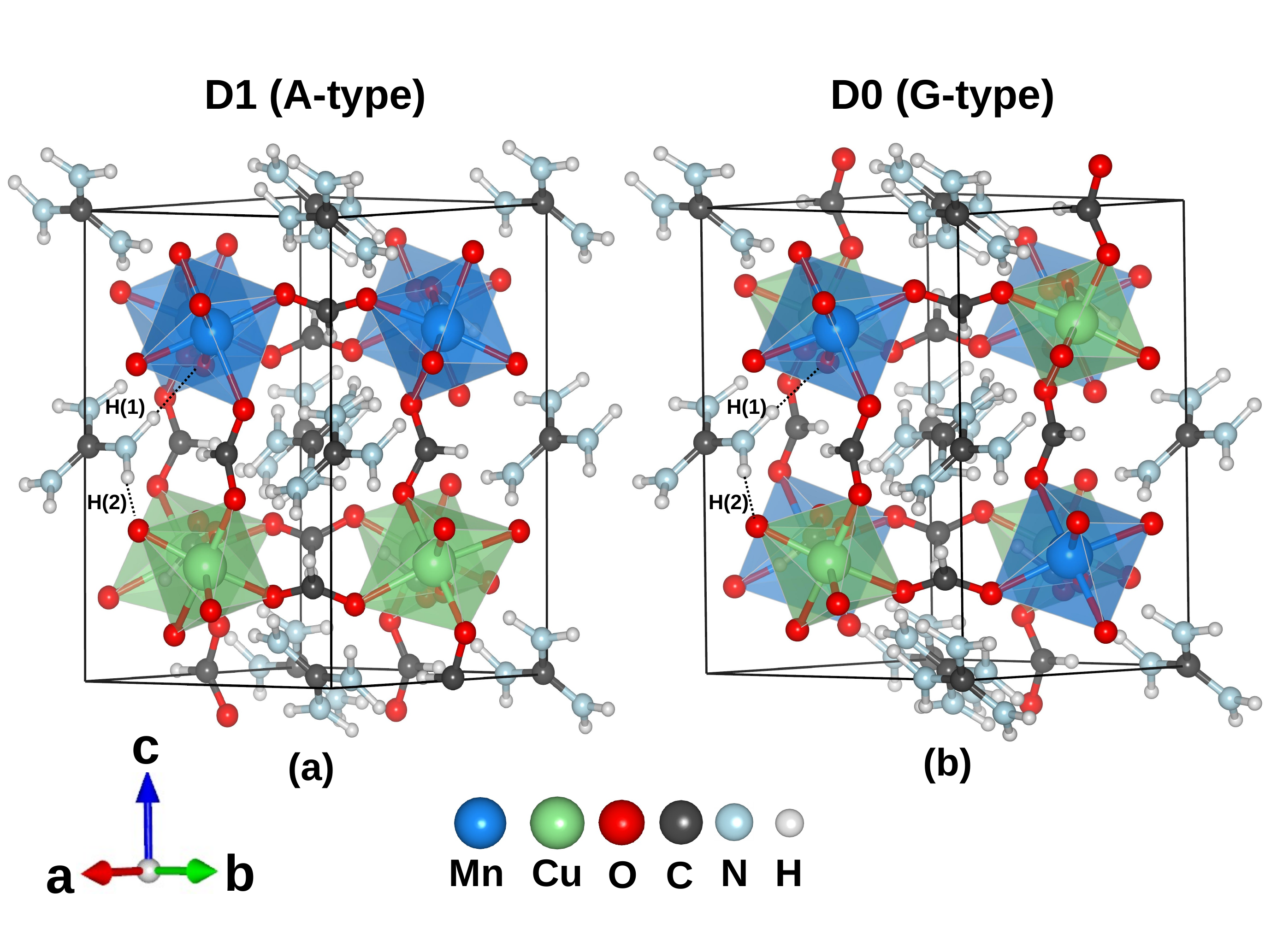}
\caption{Ball-and-stick model of two feasible structures of
(Cu$_{0.5}$Mn$_{0.5}$)-MOF in different cation orderings : (a) the D1-A
structure with alternating Cu and Mn planes perpendicular to the {\it c}-axis,
and (b) the D0-G structure with rock-salt ordering of Cu and Mn. The Cu and Mn
sublattices have opposite spins in either case, and units are connected by
HCO$_{ax}$O$_{ax}$ and HCO$_{eq}$O$_{eq}$ units in the axial and equatorial
directions, respectively. The two dashed line shows the displacement of
NH$_{2}$ group of A-site cation forming two unequal (H(1)...O$_{eq}$ and
H(2)...O$_{eq}$) bonds with the Cu and Mn octahedra sites, partially responsible
for A-site polarization. 
\label{fig:1} }
\end{figure}
Figure~\ref{fig:1}(a) shows the structure of the (Cu$_{0.5}$Mn$_{0.5}$)-MOF in
the ground-state. Each TM ion in the MOF is surrounded by six \ce{HCOO-} anions
forming a distorted octahedra. The near cubic cavities are occupied with
\ce{[C(NH2)3]+} groups providing charge balance in the compound. Like the parent
Cu-MOF, each distorted Cu-O octahedron possesses two short ( 2.02, 1.99) and
two long (2.44, 2.36) equatorial Cu-O$_{eq}$ bonds; and two medium
(2.04, 2.02) axial Cu-O$_{ax}$ bonds. The Mn-O octahedra, with two long
(2.22, 2.22) and two short (2.17, 2.21) equatorial bonds, are only slightly
distorted. Thus, the Mn-O octahedra in D1-A closely resemble those in the parent
Mn-MOF which crystallizes in a non-polar {\it Pnna} space-group.  
In the case of D0-G (Figure~\ref{fig:1}(b)), the bond-length variation around
the Cu is the same as in D1-A. However, unlike in D1-A, the octahedra around Mn
are strongly distorted with two short (2.03, 2.06), two long (2.18, 2.23) and
two medium (2.08, 2.10) bonds. This is in contrast with the parent Mn-MOF where
octahedral distortions arise only when the A site cation is
changed~\cite{Wang}. Thus, compared to D0-G,  D1-A is more stable since its
layered structure allows the Mn-O octahedra to retain the undistorted structure
seen in the parent.

\begin{figure}[pbht!]
\begin{center}
\subfigure[]
{
\includegraphics[trim=0cm 1.0cm 0.0cm
2.5cm,clip=true,width=0.5\textwidth]{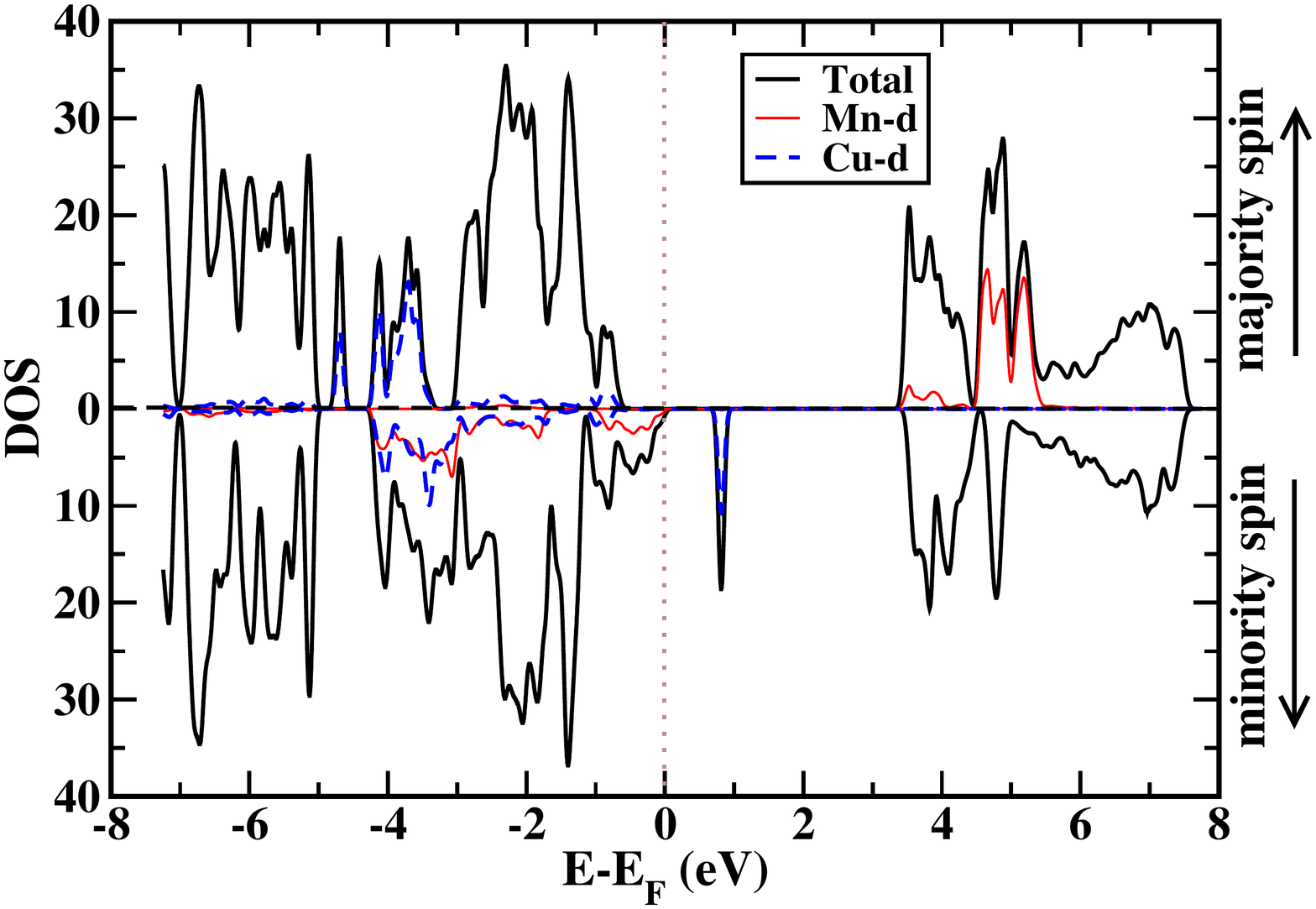}
\label{fig4:subfig1}
}
\subfigure[]
{
\includegraphics[trim=0cm 1.0cm 0.0cm
2.0cm,clip=true,width=0.5\textwidth]{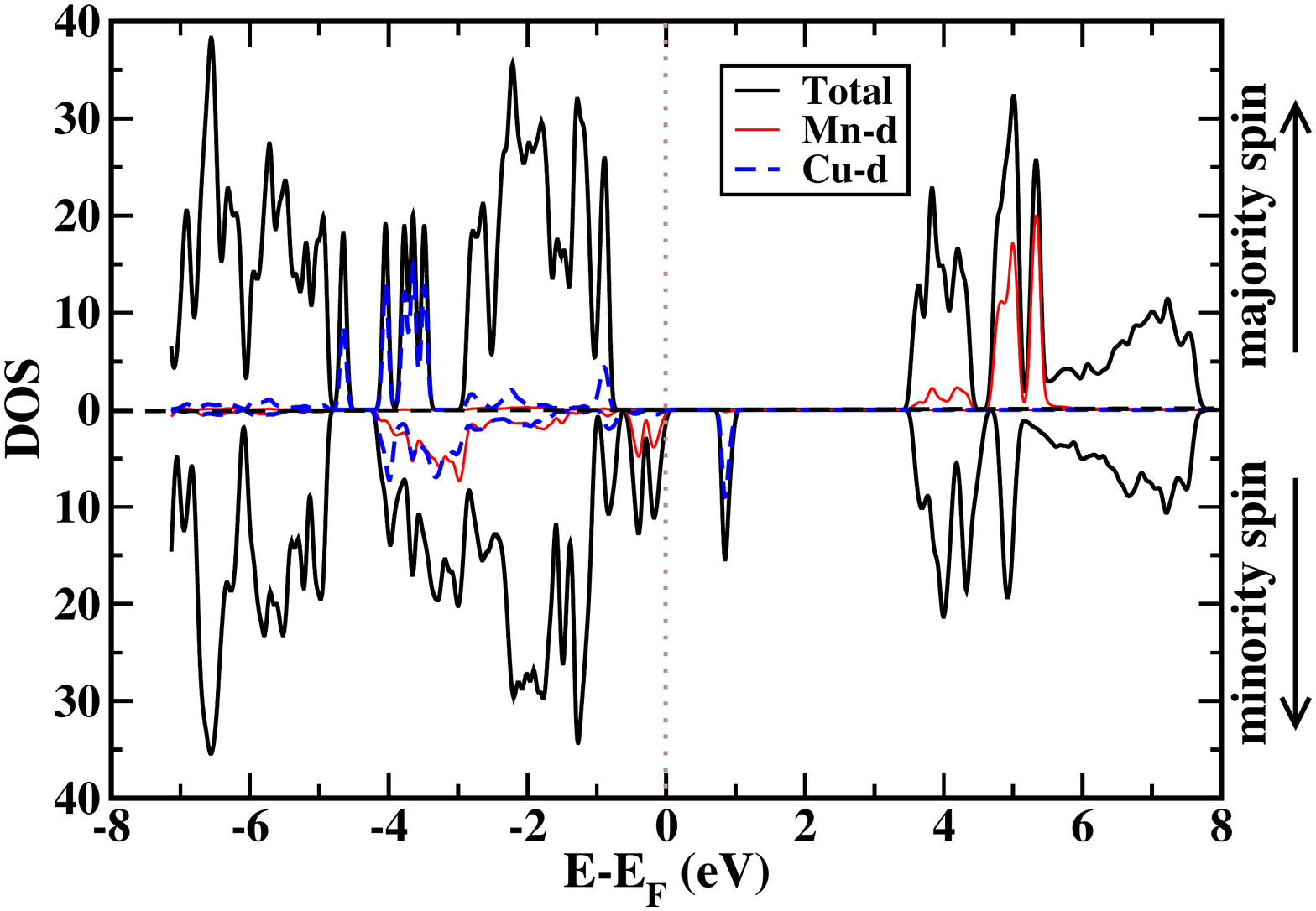}
\label{fig4:subfig2}
}
\caption{Total and projected density of states for the D1-A  and D0-G structure
in $Pna2_{1}$ phase :~(a) The GGA+$U$+vdW predicted total and projected DOS of
D1-A structure. Outer lines shows the total DOS and the inner solid lines
indicates the total $d$-orbital contribution of Mn atoms and the dashed lines
shows the $d$-orbital contribution of Cu atoms.~(b) Total and projected density
of states for the polar D0-G structure.}
\label{fig:4}
\end{center}
\end{figure}

The magnetic TM ions in the structure, linked by formate groups, interact with
each other through long-distance super-exchange~\cite{Tian} mechanism. The
density-of-states (DOS) plots for both D0 and D1 structure (see
Figure~\ref{fig:4}) show Mn to be in the high-spin Mn$^{2+}$ ($d^5$) and Cu to be
in the Cu$^{2+}$ ($d^9$) valence configurations. The valence configurations are
also confirmed by the $d$-projected occupation numbers (not mentioned here). In
both cases, the hole state from Cu forms a narrow band indicating spatial
localization. Figure~\ref{fig:4} clearly shows that the D1-A and D0-G MOFs are
ferrimagnetic insulators with a narrow band gap of 0.8 and 0.9 eV, respectively.
The DOS also reflects the AFM ordering in the structure. Partial cancellation of
moments between the two TM ions leads to a net magnetic moment of 4$\mu_B$ per
Cu-Mn pair (or 2 $\mu_B$/TM) in both D1-A and D0-G. The predicted
value is comparable to inorganic ferromagnets and higher than those generally
seen in magnetic MOFs. In D1-A, the FM interaction in the Cu layer arises due to
the AFO ordering of the Cu $d$-orbitals caused by the Jahn-Teller (JT)
effect~\cite{Stroppa}. 
As a result, the hole in Cu alternates between the two $e_g$ orbitals from one
Cu to its nearest neighbour, an effect also seen in the parent Cu-MOF.  FM
super-exchange interaction is mediated via a half-filled $e_g$ orbital on a Cu
and a completely filled one on its neighbour in the same layer, as 
predicted by the Goodenough-Kanamori (GK)~\cite{GK} rules. Surprisingly,
however, the Mn layer also displays FM order defying the GK rules for a $d^5 -
d^5$ TM ion pair. Moreover, in the D0-G case, despite the octahedral 
distortions around both TM ions, only AFM interactions prevail.

The predicted magnetic states for D0 and D1 structures can be rationalized with
the help of the exchange coupling constants for all TM pairs in the structures. These parameters 
can be extracted by mapping the DFT computed energies of the various magnetic configurations to
a nearest-neighbour ($nn$) Heisenberg Hamiltonian~\cite{paresh}. The corresponding Hamiltonians for the D0 
and D1 supercells along with a detailed description of the method of extracting the coupling constants is
presented in the SI. Taking the $ab$-plane as reference, the D0 structure has an inter- ($J^{Cu-Mn}_\perp$)
and an intra-plane ($J^{Cu-Mn}_\parallel$) Cu-Mn coupling constants.  These were calculated to be $J^{Cu-Mn}_\perp\approx$ 4.6~meV and
$J^{Cu-Mn}_\parallel\approx$ 2.53~meV, respectively. 
In D1-A, there are two in-plane ($J_\parallel^{Cu-Cu}\approx$ -0.9~meV, $J_\parallel^{Mn-Mn}\approx$
-0.5~meV) and one out-of-plane ($J_\perp\approx$ 3.9~meV) coupling constants. We
note that the Cu-Mn interactions are strongly AFM, consistent with the GK rules
for a $d^5-d^9$ pair. Thus we get a G-type AFM ordering for the D0 structure
irrespective of the JT distortions around Cu. 

In D1, the strong out-of-plane AFM
exchange along with the AFO-driven FM ordering in the Cu layer, drives the Mn
layer to be FM. This leads to the predicted A-type AFM ground state. The FM
coupling between Cu ions is key to establishing such a ground-state. To confirm this surprising result, 
we extended the range of coupling in the model used for the D1 structure to the next-nearest neighbour ($nnn$) Cu-Mn
interactions and recomputed the coupling constants. This did not affect the Mn-Mn FM coupling much (-0.52~meV) 
but instead significantly enhanced the Cu-Cu FM coupling to $J_\parallel^{Cu-Cu}\approx$ -1.71~meV. 

\begin{figure}[pbht!]
\centering
\subfigure[]
{
\includegraphics[trim=0cm 0.0cm 0cm
2.0cm,clip=true,width=0.5\textwidth]{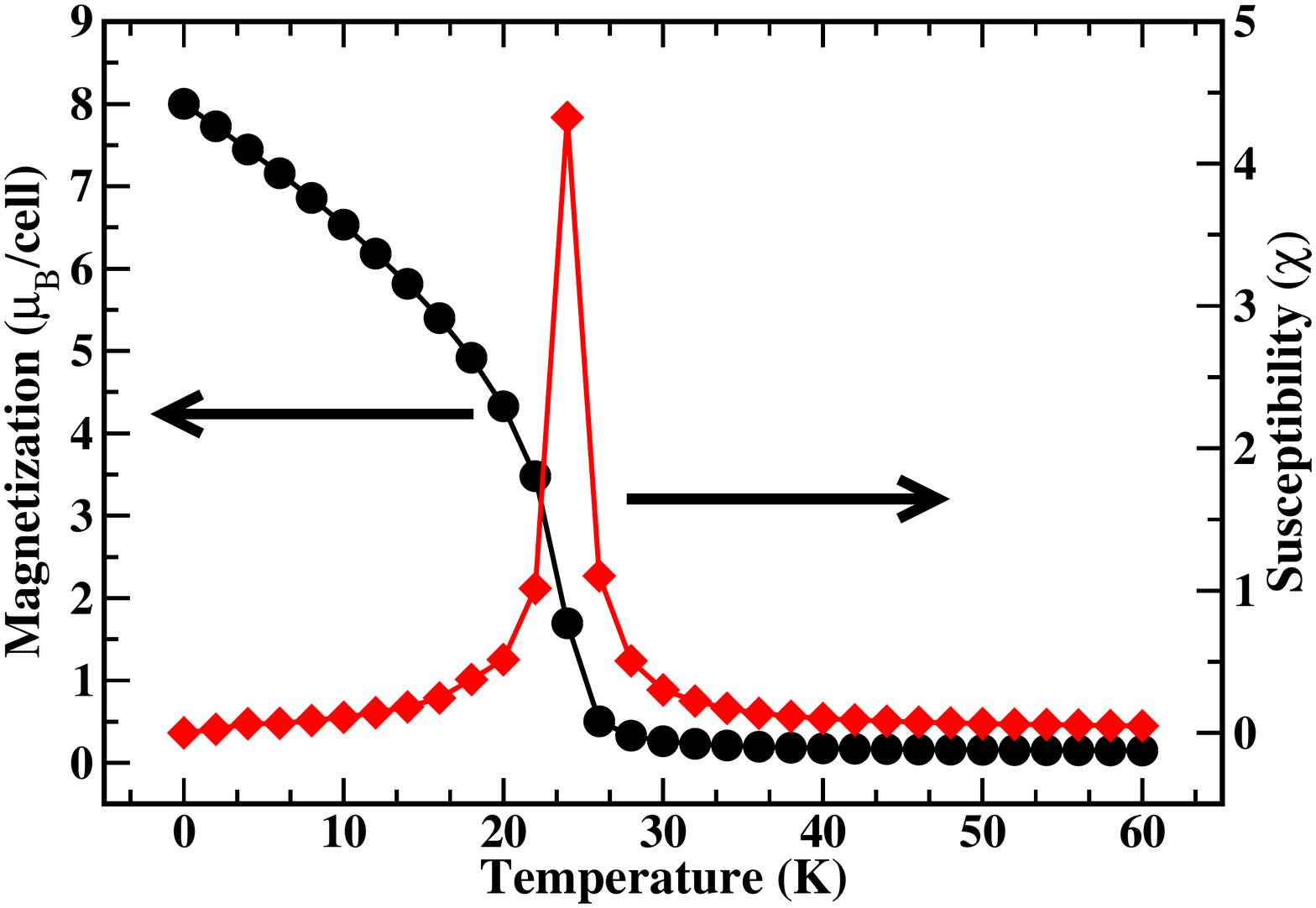}
\label{fig3:subfig1}
}
\subfigure[]
{
\includegraphics[trim=0cm 0.0cm 0cm
3.0cm,clip=true,width=0.5\textwidth]{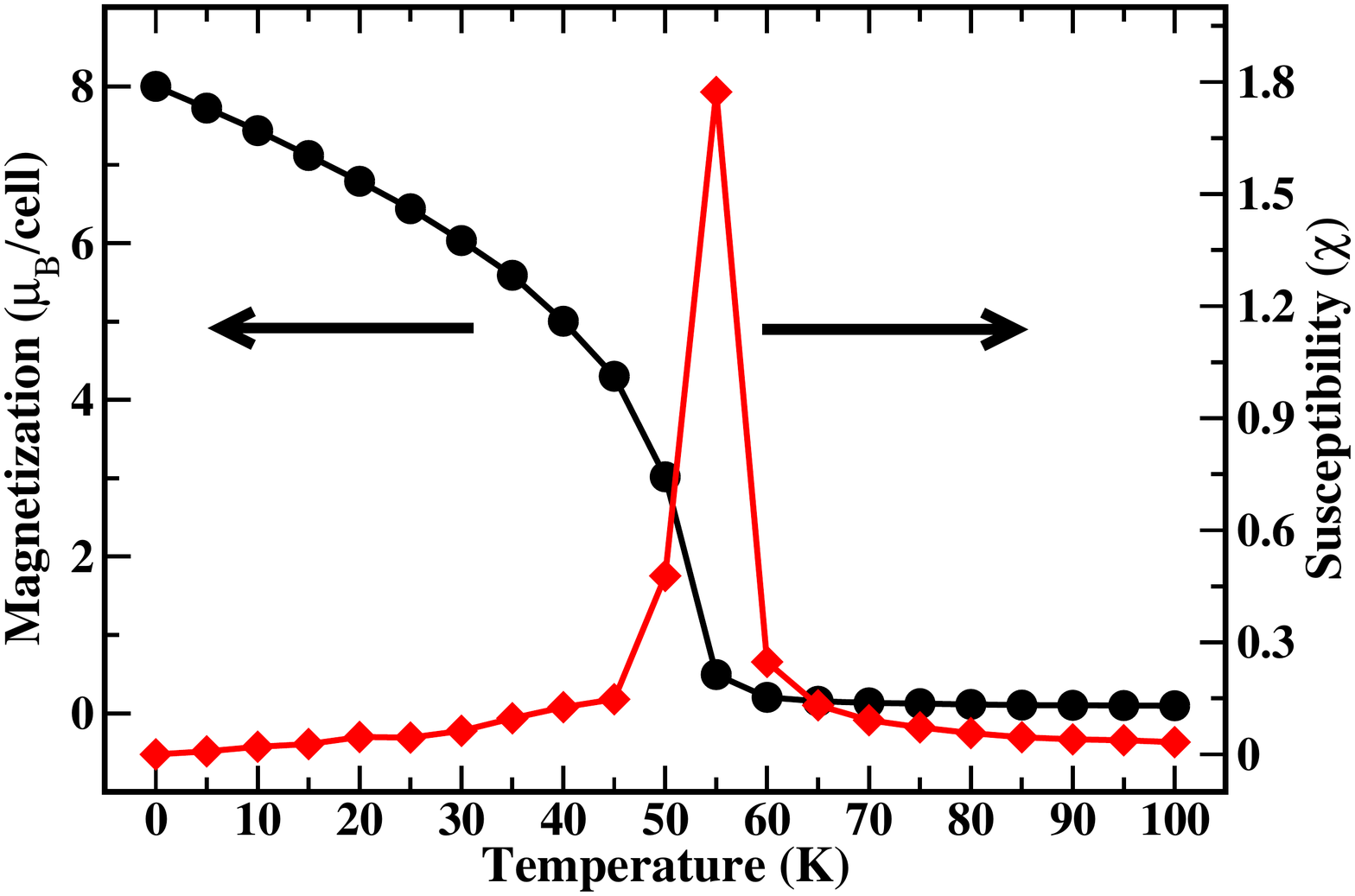}
\label{fig3:subfig2}
}
\caption{Temperature-dependence of magnetic susceptibility and 
total magnetization obtained from classical Monte Carlo simulations on (a) the
D1-A structure (the ground-state), and (b) the D0-G structure. The peak
positions of the susceptibility curves indicate that the ferrimagnetic curie
temperature ($T_{c}$) for  D1-A is 24 K and for D0-G is 56 K. In both (a) and
(b) the total magnetization rapidly increases near $T_{c}$ indicating a
paramagnetic to ferrimagnetic phase transition.
\label{fig:3}}
\end{figure}

From the magnitude of the coupling constants we anticipated a significant
increase in the magnetic transition temperature ($T_c$) as the coupling constant
is directly proportional~\cite{Fan} to $T_c$. Using classical Monte-Carlo
simulations (see SI for details), we can predict the $T_c$ for the {\it
Pna2}$_{1}$-like phase of (Cu$_{0.5}$Mn$_{0.5}$)-MOF. Figure~\ref{fig:3} shows
the magnetic moment as well as the magnetic susceptibility plotted as a function
of temperature for both structures. The plots indicate that the magnetic
transition occurs at 24 K and 56 K for D1-A and D0-G, respectively. Thus the
$T_c$ could be pushed up to $56$ K through this mixed metal strategy. The
predicted $T_c$ is a remarkable increase over that of the parent compound and is
indicative of the enhanced stability of the ferrimagnetic phase in the
(Cu$_{0.5}$Mn$_{0.5}$)-MOF relative to most other magnetic MOFs seen so far. The
estimates given here are based on the $nn$ Heisenberg model. In the case of D1, use 
of the coupling constants based on the $nnn$ model yielded a $T_c = 38$ K (see SI).

First-principles calculations on Cu-MOF have estimated a {\it c}-axis electric
polarization of 0.37 $\mu {\rm C}/{\rm cm}^2$~\cite{Stroppa}, while Mn-MOF was
found to crystallize in a non-polar structure~\cite{Hu}. It has been suggested
that the weak polarization can be tuned by varying the organic A-site
cation~\cite{Sante} or by strain field~\cite{Ghosh}. Indeed,
[\ce{CH3CH2NH3][Mn(HCOO)3}] was found to yield a theoretical polarization of 1.6
$\mu$C/cm$^{2}$~\cite{Sante} with some contribution arising from octahedral
distortion around Mn cations. While the B-site mixing strategy proposed here was
aimed mainly at improving the magnetic moments, we also investigated the
polarization of the predicted compounds. We calculated the electric polarization
using a Berry phase approach~\cite{king} ensuring the convergence of the
computed numbers with the relevant parameters (see SI). Surprisingly, we found
that both D0-G and D1-A yielded a significantly enhanced {\it c}-axis
polarization of -9.93 and -9.77 $\mu$C/cm$^{2}$, respectively, than that 
(0.37 $\mu {\rm C}/{\rm cm}^2$) in the parent Cu-MOF. We obtained the
polarization as a difference between the polar ($\lambda=1$) and non-polar
($\lambda=0$) structures. Note that the polarization difference
computed this way is one value in a lattice of values spaced by the polarization
quantum. However, the actual value can be fixed by looking at the changes in the
Berry phase along a smooth path connecting the polar and non-polar structures.
\begin{figure}[t!]
\centering
\subfigure[]
{
\includegraphics[trim=0cm 0.0cm 0cm
3.0cm,clip=true,width=0.5\textwidth, clip=]{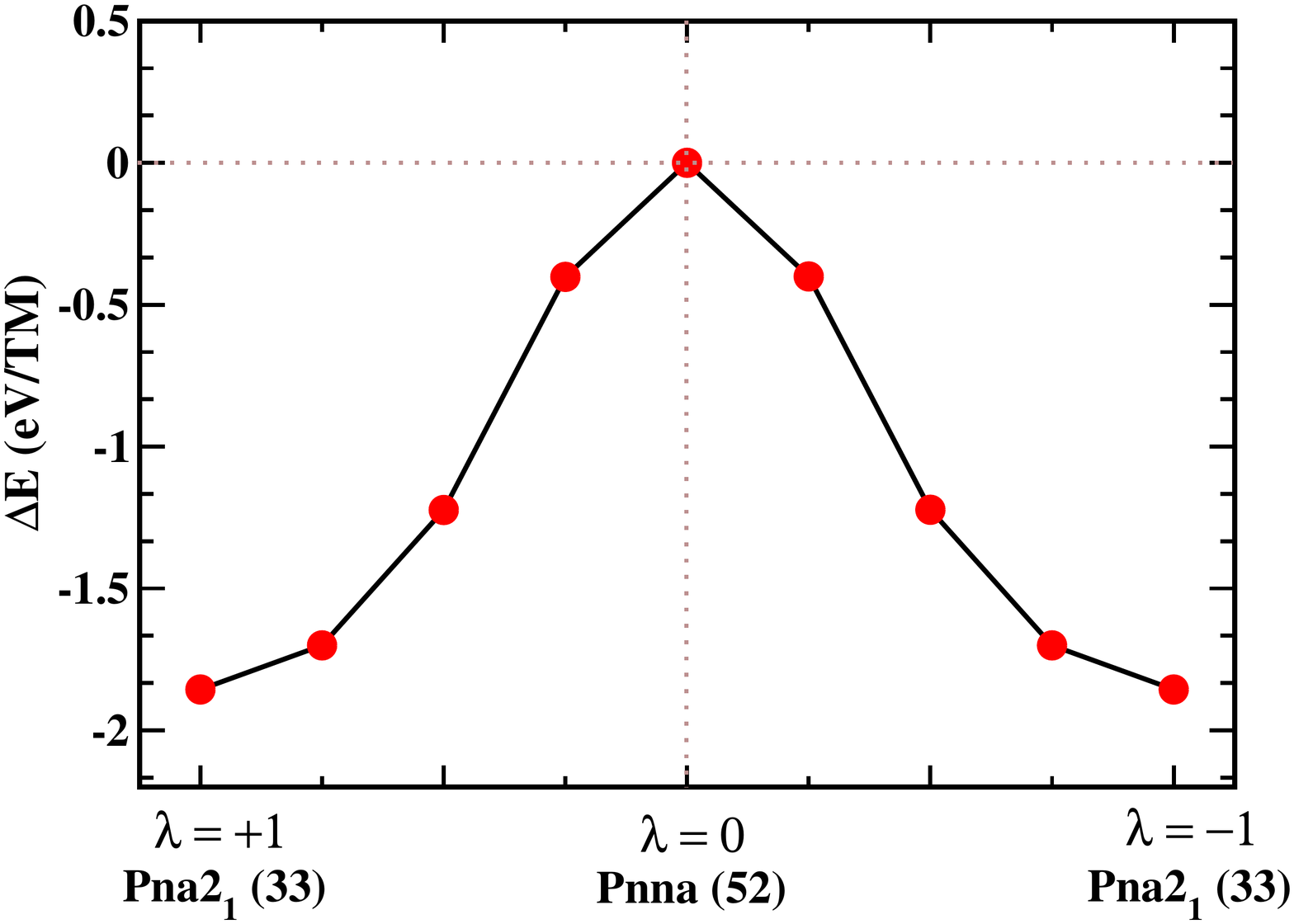}
\label{fig5:subfig1}
}
\subfigure[]
{
\includegraphics[trim=0cm 0.0cm 0cm
3.0cm,clip=true, width=0.5\textwidth]{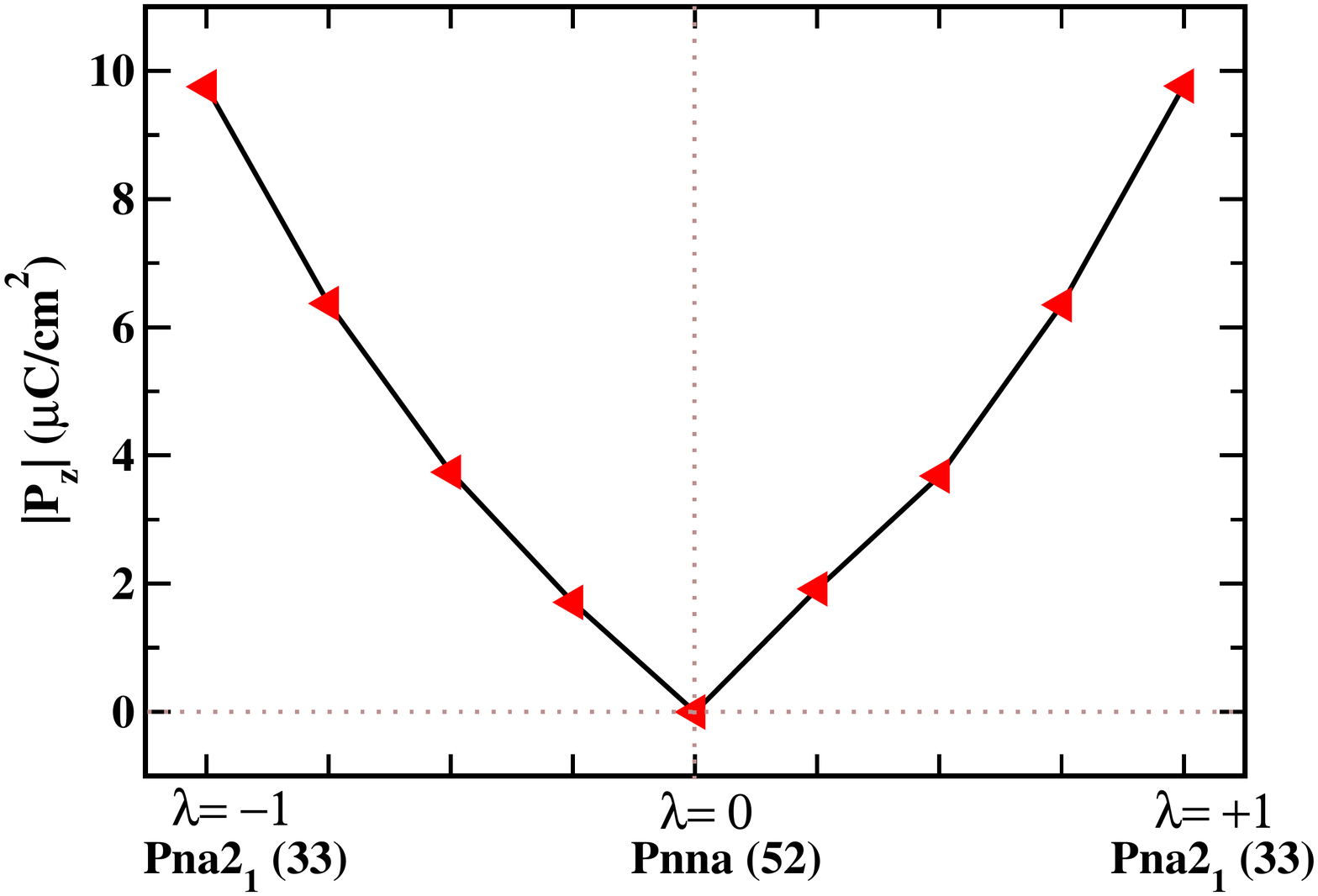}
\label{fig5:subfig2}
}
\caption{(a)~The variation of total energy difference as a function of the
stuctural distortion from paraelectric to the polar D1-A structure. (b)
Variation of
total ferroelectric polarization in D1-A along c-axis as a function of amplitude
of
polar distortion ($\lambda$).\label{fig:5}} 
\end{figure}
We constructed various structures linearly interpolated between polar ({\it
Pna2$_1$-like}) and the paraelectric phases. The latter was assumed to be the
non-polar  {\it Pnna}-like centro-symmetric structure~
\cite{aroyo,capi}. The structures along the interpolation were followed using a
parameter $\lambda$, measuring the amplitude of the displacement, with values
$\pm 1$ for the polar and $0$ for the non-polar centric forms, respectively.
Figure~\ref{fig:5} depicts
the interpolation thus done in the D1-A structure. The maximum atomic
displacement between $\lambda=0$ and
$\lvert\lambda\rvert=1$ was found to be about $0.26~{\rm \AA}$. The variation in
energy of D1-A  along an idealized polarization switching path through the
non-polar intermediate is shown in Figure~\ref{fig:5}(a).
The polar phase of D1-A is more stable that the centric phase by $1.9$ eV/TM. In
Figure~\ref{fig:5}(b), we have plotted the ferroelectric polarization $P_{z}$
along the polar $c$-axis as a function of $\lambda$. The polarization calculated
from this plot is found to be $-9.77~\mu {\rm C/cm}^2$. A similar approach was 
also followed for the D0-G structure resulting in an energy difference of $1.5$
eV/TM between the polar and non-polar phases (see Figure in SI). A polarization
of $-9.93~\mu {\rm C/cm}^2$ was computed in this case from a 2-point Berry phase formula.

The large energy differences between the polar and centric structures noted
above suggest insurmountable ferroelectric switching barriers.
However, this interpretation is not always warranted. While we have considered
an idealized path for the ferroelectric switching, this may not necessarily be
the path followed by the system in reality. For instance, the barrier for the
ground state D1-A structure is reduced to 1.2 eV, when we optimize
the cell parameters of the \textit{Pnna} structure keeping the ions in their
centro-symmetric positions. The resulting orthorhombic unit cell (see SI) hints
at a structural phase transition accompanying the polarization switching.
Furthermore, the barrier calculated here is a single domain switching barrier.
In practice, however, the ferroelectric switching barrier can be significantly
lowered by presence of domains~\cite{Yoon, Levanyuk} not considered in the
present calculation. Indeed, recently, Somdutta \textit{et al.} have
experimentally shown ferroelectric switching in \ce{GaFeO3} thin-films although
the bulk form was theoretically predicted to have high polarization switching
barrier~\cite{Somdutta}. They have attributed the reduction of ferroelectric
barriers to the presence of ferroelectric domains in these samples. The higher
the nucleation of domains the more the reduction of thermally insurmountable
single-domain switching barrier~\cite{Yoon}. Similar effects could, in
principle, also lower the switching barriers in the predicted MOF which can be
verified through experimental realization of the system.

In Cu-MOF, it was shown~\cite{Stroppa} that the displacements of \ce{NH2} groups
of the guanidinium cations result in the dominant contribution to the
ferroelectric polarization. In contrast, in the ground-state of the
(Cu$_{0.5}$Mn$_{0.5}$)-MOF, we found that the larger contribution arises from
the \ce{BX3} group instead of A-site. In order to estimate their relative
magnitudes, we calculated the polarization arising from A-site displacements
({\it P}$_A$) and displacement of atoms belonging to the functional group
\ce{BX3} ({\it P}$_{BX_3}$) separately by displacing each group towards its
polar configuration
keeping the others fixed in the non-polar geometry. We found the values {\it
P}$_A =$ 0.21 and {\it P}$_{BX_3}= $ -7.36 $\mu {\rm C}/{\rm cm}^{2}$,
respectively, indicating that the major contribution is made by the
distortions at the \ce{BX3}-site. The estimated polarization value of A-site
cation ion is in excellent agreement with the previously estimated value (0.21) of
the parent Cu-MOF~\cite{Stroppa}. The significantly larger polarization
arising from the \ce{BX3} framework in this case is clearly due to the presence
of the non-JT Mn ions which were absent in the parent Cu-MOF. The discrepancy
between the polarization estimated from these contributions (-7.2 $\mu {\rm
C}/{\rm cm}^{2}$) and the exact value is likely due to the neglect of relaxation
effects in the former. 

In order to further support our Berry phase results, we have calculated the
ferroelectric polarization from the dipole moments contributed from each
A-site and \ce{BX3} group by using a non-periodic, localized basis
code~\cite{G09} (see SI Table VI for details). We found the polarization values
to be {\it
P}$_A =$ 0.18 and {\it P}$_{BX_3}=$-5.13 $\mu {\rm C}/{\rm cm}^{2}$,
respectively. These values mirror the contributions seen in the Berry phase
approach and once again confirm that the major contribution to
the total polarization arises from the distortion of \ce{BX3} groups. We also 
applied the method to estimate the contributions to the polarization in the
Cu-MOF. The results (see SI) confirmed that the dipole moments arising from the
\ce{BX3} groups significantly increase in the mixed metal MOF compared to the
parent Cu-MOF while there was no change in the moments at the A-site.

The existence of magnetoelectric coupling has been demonstrated in the parent MOF and
is an important ingredient for applicability of these materials. In order to test for the coupling in the 
mixed metal MOF, we also performed DFT+$U$+vdW calculations incorporating
spin-orbit coupling on the D1-A and D0-G structures. The resulting magnetization
values
are summarized in Table~\ref{tbl:7}. The magnetization along the $z$-axis was
not affected (4.0 $\mu_B$ per TM pair) but it developed components in the
$xy$-plane of magnitude 0.01 $\mu_B$ per TM pair. Inverting the direction of the
polarization leads to retention of the $z$-axis component but inversion of the
component in the $xy$-plane. So there is indeed a magneto-electric effect
confined to the $xy$-plane similar to the case of the parent MOF. The magnitude
of the moments are slightly reduced by mixing in Mn ions as it is the JT-active
Cu ions in the system which mostly contribute to the ME coupling (see
Table V-VIII in SI). Thus, the mixing in of non-JT ions leads to an apparent
suppression of the magnetoelectric effect. 
\begin{table}[h!]
 \caption{Calculated magnetic moments ($\mu_B/cell$) for D1 and D0 structures.
}
  \begin{ruledtabular}
\begin{tabular}{@{\hspace{3em}} l l c  @{\hspace{3em}}}
& ~~~~~~~~~~~\textrm{D1-A type} & ~~~~\textrm{D0-G type} \\
\colrule   
  Distortion  &~~~~~~m$_x$~~~~~~m$_y$~~~~~~m$_z$~~~ & ~~~~m$_x$~~~~~
m$_y$~~~~~ m$_z$ \\
\hline
 $\lambda = +1$  & ~~~-0.02~~~~0.00~~~~-8.01&~~~-0.03~~~~-0.01~~~~-8.01 \\
 $\lambda = -1$  & ~~~~0.01~~~~0.01~~~~-8.01&~~~~0.00~~~~~0.01~~~~-8.01\\ 
\end{tabular}
\end{ruledtabular}
\label{tbl:7}
\end{table}

In conclusion, we have designed, from first-principles, a mixed metal
perovskite MOF, [{C(NH$_{2}$)$_{3}$][(Cu$_{0.5}$Mn$_{0.5}$)(HCOO)$_{3}$] with
significantly enhanced magnetization and a polarization compared to its parent
Cu-MOF as well as other mixed metal MOFs synthesized so far~\cite{Mazaka}. We
also predict that the ground state MOF would have a magnetic
transition temperature of around 24 K which can be enhanced up to 56 K by
altering the cation ordering in the B-site. This is a remarkable improvement
over multiferroic MOFs synthesized so far. Our calculations indicate large
formation enthalpies for the compound in two lowest energy structures suggesting
feasibility of laboratory synthesis. The ground-state structure is composed of
layers of Mn and Cu alternating along the {\it c}-axis. A strong AFM Cu-Mn
exchange coupling along with FM ordering in the Cu layer, driven by Jahn-Teller
distortion, forces FM coupling in the Mn layer as well. This results in an
A-type AFM ordered state with a magnetic moment of 2 $\mu_B$/TM. Changes in
hydrogen-bonds at the A-site, distortions of the oxygen octahedra around Cu and
Mn, as well as displacements of the formates contribute to the polarization
enhancement. The competing magnetic interactions between the Cu and Mn layers
suggest the possibility of magnetic and structural transitions with variation of
relative composition~\cite{CuCd} of the two TM ions as well as epitaxial strain.
These will be the subjects of a future study. 

Our choice of the TM ions as well as the feasibility of the mixed metal approach
are motivated by the facts that, (i) polar [\ce{C(NH2)3][Cu(HCOO)3}] and
[\ce{CH3CH2NH3][Mn(HCOO)3}] have already been experimentally synthesized, and
(ii) very recently~\cite{CuCd}, a mixed metal MOF with the same framework has
been synthesized. Therefore, we expect that
[{C(NH$_{2}$)$_{3}$][(Cu$_{0.5}$Mn$_{0.5}$)(HCOO)$_{3}$] can also be realized.
The strategy can be used to further explore other metal combinations in the
($A_{2}BB'X_{6}$) structure, along with variations in their compositions, to
engineer the magnetic, electric~\cite{CuCd} and even elastic~\cite{Cheetam}
properties in this class of MOFs. 

\acknowledgements{The authors would like to thank all the members of AIT group
of IISER Bhopal for valuable discussions. The authors gratefully acknowledge
Indian Institute of Science Education and Research Bhopal HPC facility for
computational resources. P.C.R. would like to acknowledge CSIR-HRDG for funding
through CSIR-JRF programme.}

\bibliographystyle{apsrev4-1}
\bibliography{./reference}
\end{document}